\documentclass{ws-procs9x6}

\def\nc{{N_c}}
\def\ga{{g_A}}
\def\yop{{f_\pi}}
\def\yo2{{f_\pi^2}}
\def\sss{\scriptscriptstyle}

\begin{document}

\title{Exploring Baryon Chiral Multiplets}

\author{S.R.~Beane}

\address{Institute for Nuclear Theory,\\
University of Washington,\\
Seattle, WA 98195-1550\\ 
E-mail: sbeane@phys.washington.edu}


\maketitle

\abstracts{The full QCD chiral symmetry algebra has predictive consequences at
  low energies. I discuss the ground-state chiral multiplet involving the light
  baryons and emphasize the special role of the Roper resonance.}

\section{Introduction}
\label{sec:intro}

The full QCD chiral symmetry group
---$SU(2)\times SU(2)$ in the case of two massless flavors--- has
important algebraic consequences at low energies. There is a sense
in which hadrons fall into (generally-reducible) representations of $SU(2)\times SU(2)$ {\it for each
  helicity}~\cite{Fred,Weinberga}. All of the consequences of chiral symmetry
for hadron masses and pion transition amplitudes can be found by considering
sum rules derived from unsubtracted dispersion relations and saturated by
single-particle states~\cite{Fred}. However, the symmetry interpretation is more powerful
and intuitive and allows one to discuss hadrons in the language of the
underlying theory~\cite{Weinberga}. The ground-state chiral representation for the light
baryons, which involves the nucleon and its chiral partners is, of course, of
fundamental interest and is the subject of this talk. In order to get a sense
of the particle content of this representation, I will first discuss an updated
analysis of the well-known Adler-Weisberger (A-W) sum rule for pion-nucleon
($\pi N$) scattering~\cite{adler}. Remarkably, this sum rule suggests that the
nucleon chiral representation ---to some degree of approximation--- involves
only a few states. While this is partly understood as a consequence of the
large-$\nc$ approximation~\cite{largeN,Weinbergc,Beane4}, what I will describe
here is different in a subtle way; in particular, the A-W sum rule suggests
that the nucleon, $N$, is joined by $\Delta$ {\it and} by the Roper, $N^\prime$, in
a chiral representation.  Armed with this information, I will then
construct this reducible chiral representation 
using tensor analysis and show that the symmetry information is
equivalent to the complete set of A-W sum rules for $\pi B$ scattering (both
elastic and inelastic) where $B$ is a baryon in the ground state multiplet. I
will also discuss the chiral transformation properties of the baryon
mass-squared matrix and their consequences. An interpretation of
the ground state chiral multiplet in the context of the naive constituent quark
model (NCQM) is then offered. Finally, I conclude.

\section{Adler-Weisberger Sum Rules}
\label{sec:AW}

Consider the renowned Adler-Weisberger sum rule~\cite{adler},

\begin{equation}
g_A^2=1-{{2\yo2}\over\pi}\int_0^\infty {{d\nu}\over\nu}
\lbrack\sigma^{\pi^- p}(\nu )-\sigma^{\pi^+ p}(\nu )\rbrack .
\label{eq:AW}
\end{equation}
Here $\ga$ is the nucleon axial-vector coupling, ${\yop}\simeq 93~{\rm MeV}$ is
the pion decay constant and $\sigma^{\pi^\pm p}$ is the total cross-section for
charged pion scattering on a proton.
Recall that this sum rule for the $\pi N$ scattering amplitude follows from two
inputs: (i) a chiral symmetry low-energy theorem and (ii) the assumption that
the forward $\pi N$ amplitude with isospin, $I=1$, in the $t$-channel satisfies 
an unsubtracted dispersion relation. Saturating the sum rule with $N$ ($I=1/2$) and $\Delta$
($I=3/2$) resonances gives

\begin{equation}
g_A^2=1-\sum_{\sss N} {\bf I}_{\sss N}+\sum_{\sss \Delta} {\bf I}_{\sss \Delta}
+\, {\it continuum},
\label{eq:AWRS}
\end{equation}
where the ${\bf I}_{R}$ are related to experimental widths by

\begin{equation}
{\bf I}_{R} ={{{64\pi\yo2}{M_{R}^3}}\over{3(M_{R}^2-M_N^2)^3}}
\left(S_{R}+\frac{1}{2}\right)
\Gamma^{\scriptstyle {\rm TOT}}({R}\rightarrow N\pi ),
\label{eq:RSdef}
\end{equation}
and $S_{R}$ is the spin of the resonance ${R}$.  

We can now go to the Particle Data Group (PDG)~\cite{pdg02} and compute the
contribution of each $N$ and $\Delta$ state to the sum rule (see Table~\ref{tab1}).  We
include only established resonances ($\star\star\star$ and
$\star\star\star\;\star$), using PDG central values and estimates. We find
$\sum {\bf I}_N=0.72$ and $\sum {\bf I}_\Delta =1.3$.  Neglecting the
continuum contribution (we will return to this point below), we then obtain
$\ga =1.26$, to be compared to the experimental value of $1.2670\pm 0.0035$~\cite{pdg02}. 
This is truly remarkable agreement.  There are several
important things to notice from Table~\ref{tab1}.  First, there is a cancellation
between the $N$- and $\Delta$-type contributions, which enter with opposite
sign.  Second, $\Delta (1232)$ and $N(1440)$ dominate the sum rule.  Axial
transitions of the excited baryons to the ground-state nucleon are small
compared to the dominant transitions.  For instance, saturating the sum rule
with these two states alone gives $\ga=1.34$.

\begin{table}[htb]
\tbl{Resonances which contribute to the A-W sum rule for $\pi N$ scattering. 
We have used PDG central values and estimates. We emphasize that there is substantial uncertainty in these values.
Only established resonances ($\star\star\star$ and $\star\star\star\;\star$) have been tabulated.\vspace*{1pt}}
{\footnotesize
\begin{tabular}{|c|c|c||c|c|c|}
\hline
{}    & $R$  & ${\bf I}_R$ & ${}$  & $R$ & ${\bf I}_R$  \\
\hline
$P_{11}\;({1\over 2}^+)$ & $N(940)$  &  $\>\;\; --$ & $P_{11}\;({1\over 2}^+)$  &  $N(1710)$  &  $\>\;\;0.01$ \\
$P_{33}\;({3\over 2}^+)$ & $\Delta (1232)$  &  $\>\;\;1.02$ & $P_{13}\;({3\over 2}^+)$  &  $N(1720)$  &  $\>\;\;0.02$ \\
$P_{11}\;({1\over 2}^+)$ & $N(1440)$   &  $\>\;\;0.23$ & $F_{35}\;({5\over 2}^+)$  &  $\Delta (1905)$  &  $\>\;\;0.02$ \\
$D_{13}\;({3\over 2}^-)$ & $N(1520)$   &  $\>\;\;0.09$ & $P_{31}\;({1\over 2}^+)$  &  $\Delta (1910)$  &  $\>\;\;0.01$ \\
$S_{11}\;({1\over 2}^-)$ & $N(1535)$   &  $\>\;\;0.04$ & $P_{33}\;({3\over 2}^+)$  &  $\Delta (1920)$  &  $\>\;\;0.01$ \\
$P_{33}\;({3\over 2}^+)$ & $\Delta (1600)$   &  $\>\;\;0.06$ & $D_{35}\;({5\over 2}^-)$ &  $\Delta (1930)$  &  $\>\;\;0.03$ \\
$S_{31}\;({1\over 2}^-)$ & $\Delta (1620)$   &  $\>\;\;0.02$ & $F_{37}\;({7\over 2}^+)$ &  $\Delta (1950)$  &  $\>\;\;0.08$ \\
$S_{11}\;({1\over 2}^-)$ & $N(1650)$   &  $\>\;\;0.04$ & $G_{17}\;({7\over 2}^-)$  &  $N(2190)$  &  $\>\;\;0.03$ \\
$D_{15}\;({5\over 2}^-)$ & $N(1675)$   &  $\>\;\;0.08$ & $H_{19}\;({9\over 2}^+)$  &  $N(2220)$  &  $\>\;\;0.03$ \\
$F_{15}\;({5\over 2}^+)$ & $N(1680)$   &  $\>\;\;0.10$ & $G_{19}\;({9\over 2}^-)$  &  $N(2250)$  &  $\>\;\;0.02$ \\
$D_{13}\;({3\over 2}^-)$ & $N(1700)$ & $\>\;\;0.01$ & $H_{3,11}\;({{11}\over 2}^+)$ & $\Delta (2420)$  &  $\>\;\;0.02$ \\
$D_{33}\;({3\over 2}^-)$ & $\Delta (1700)$ & $\>\;\;0.03$ & $I_{1,11}\;({{11}\over 2}^-)$  &  $N(2600)$  &  $\>\;\;0.02$ \\ 
\hline
\end{tabular}\label{tab1} }
\vspace*{-13pt}
\end{table}

Given the uncertainties in the resonance masses and axial couplings, and the
neglect of the continuum contribution, such remarkable agreement must to some
degree be fortuitous. Given the success of the sum rule one might ask: what
precisely is the sum rule testing about QCD? What is the significance of the
assumption about the asymptotic behavior of the forward $\pi N$ scattering
amplitude? Why do $\Delta (1232)$ and $N(1440)$ seem to have special status in
saturating the sum rule?  In order to answer these questions we will rephrase
the discussion of the sum rule entirely in the language of chiral symmetry.

\section{The Ground State Chiral Multiplet}
\label{sec:GS}

In the limit of vanishing up and down quark masses, QCD has an $SU(2)_L\times
SU(2)_R$ invariance. We can write the chiral algebra as

\begin{equation}
[{{\mathcal Q}^{\sss A}_{\alpha}},{{\mathcal Q}^{\sss A}_{\beta}}]=i{\epsilon_{\alpha\beta\gamma}}{T_\gamma}; 
\qquad
[{T^{\;\;}_\alpha},{{\mathcal Q}^{\sss A}_{\beta}}]=
i{\epsilon_{\alpha\beta\gamma}}{{\mathcal Q}^{\sss A}_{\gamma}};
\qquad
[{T^{\;\;}_\alpha},{T^{\;\;}_\beta}]=i{\epsilon_{\alpha\beta\gamma}}{T^{\;\;}_\gamma},
\label{eq:qcdsymm}
\end{equation}
where ${T_\alpha}$ are $SU(2)_V$ generators and ${{\mathcal Q}^{\sss A}_{\alpha}}$ are the
remaining axial generators. We define the axial-vector coupling matrix,

\begin{equation}
<{h^\prime , \lambda^\prime}|{{\mathcal Q}^{\sss A}_\alpha}|{h, \lambda}>=
[{X_\alpha^\lambda}]_{\sss{h^\prime}h}\,\delta_{\sss{\lambda\lambda'}},
\label{eq:qcdsymm2}
\end{equation}
where $|{h, \lambda}>$ is a baryon state of definite helicity
$\lambda$. Notice the Kronecker delta on the right side of this
equation.  This implies that we are defining ${X_\alpha^\lambda}$ in a
helicity-conserving Lorentz frame~\cite{Weinberga}.  A frame in which all momenta are collinear
is such a frame, as is the infinite-momentum frame. 
Taking matrix elements of the $SU(2)\times SU(2)$
algebra of Eq.~(\ref{eq:qcdsymm}) and inserting a complete set of states gives

\begin{equation}
[{X_\alpha^\lambda},{X_\beta^\lambda}]_{\sss{h^\prime}h}=
i{\epsilon_{\alpha\beta\gamma}}[{T_\gamma}]_{\sss{h^\prime}h}.
\label{eq:qcdsymm3}
\end{equation}
This is a (generalized) A-W sum rule. 
Before considering the consequences of this sum rule, we will define the axial
couplings of the nucleon, $N$, the Roper, $N^\prime$, and Delta, $\Delta$.
At leading order (LO) in chiral perturbation theory the relevant axial matrix elements are
defined through the currents~\cite{JM}
\begin{eqnarray}
&& J^{\alpha, 5}_{{\uparrow},LO}
 =  g_A\ N_\uparrow^\dagger \ T^\alpha\  N^{}_\uparrow +
g_A^\prime\ \left(\ N_\uparrow^\dagger \ T^\alpha\  N^\prime_\uparrow 
\ +\ {\rm h.c.}\ \right)
\ +\ g^{\prime\prime}_A\ N_\uparrow^{\prime\dagger} \ T^\alpha\  N^\prime_\uparrow \nonumber \\
&&\ -\ {\mathcal C}_{\Delta N} \ \left(\ \sqrt{\textstyle{2\over 3}} N_\uparrow^\dagger \ T^\alpha\  
\Delta^{}_\uparrow \ +\ {\rm h.c.}\ \right)
\ -\ {\mathcal C}_{\Delta N^\prime}\ \left(\ \sqrt{\textstyle{2\over 3}} 
N_\uparrow^{\prime\dagger} \ T^\alpha\ \Delta^{}_\uparrow 
\ +\ {\rm h.c.}\ \right) \nonumber \\
&&\ -\ {\mathcal H}_{\Delta\Delta} \textstyle{1\over 3} 
\Delta_\uparrow^\dagger\ T^\alpha\
\Delta^{}_\uparrow \ \ , \nonumber \\
&& J^{\alpha, 5}_{{\Uparrow},LO}
 =  
\ -\ {\mathcal H}_{\Delta\Delta}\ \Delta_\Uparrow^\dagger\ T^\alpha\
\Delta^{}_\Uparrow
\label{eq:chiptaxial}
\end{eqnarray}
for the $\lambda={1\over 2}$ helicity states ($\uparrow$) and
$\lambda={3\over 2}$ helicity states ($\Uparrow$), respectively.
(An aside: the NCQM places the $N$ and $\Delta$ in the $\bf{20}$-dimensional
representation of spin-flavor $SU(4)$, and
the $N^\prime$ and a $\Delta^\prime$ in the $\bf{20}^\prime$ 
representation. This leads to the familiar NCQM predictions: 
$g_A=g_A^{\prime\prime}={5\over 3}$, $g^\prime_A=0$,
${\mathcal C}_{\Delta N}=-2$ and ${\mathcal H}_{\Delta\Delta}=-3$.)
Choosing $h=h^\prime=N$ and taking $N^\prime$ and $\Delta$ as intermediate
states it immediately follows from Eq.~(\ref{eq:qcdsymm3}) that

\begin{equation}
g_A^2+g_A^{\prime\,2}=1+{\textstyle{4\over 9}}\ {\mathcal C}_{\sss\Delta N}^2 \ .
\label{eq:AWpolesat}
\end{equation}
One can easily verify that this is precisely what one obtains from the more
conventional form of the A-W sum rule given in Eq.~(\ref{eq:AWRS}).

One might worry that the vacuum should contribute in the sum over states 
in Eq.~(\ref{eq:qcdsymm3}) and
that the axial generator acting on the vacuum will generate quark-antiquark
pairs, thus destroying the algebraic structure of the sum rule. The advantage of working in
a helicity-conserving frame is that the vacuum does not contribute in the sum
over states; i.e. ${{\mathcal Q}^{\sss A}_{\alpha}}|{0}>=0$.
The chiral symmetry is, however, broken 
spontaneously: although ${X_\alpha^\lambda}$ satisfies the chiral
algebra, it does not commute with the baryon mass-squared matrix and is therefore not a
symmetry generator. Hence in helicity-conserving frames, all evidence of
symmetry breaking is in the Hamiltonian and not in the states.

In principle, the mass-squared matrix, ${\hat M}^2$, can transform as a
singlet plus any non-trivial representation(s) of the chiral group. 
In Ref.~\cite{Weinberga}, Weinberg showed that the assumption that the forward
$\pi N$ amplitude with $I=2$ in the $t$-channel satisfies a
superconvergence relation implies that

\begin{equation}
{\hat M}^2={\hat M}_{\bf 1}^2+{\hat M}_{\bf 22}^2,
\label{eq:msm1}
\end{equation}
where ${\hat M}_{\bf 1}^2$ and ${\hat M}_{\bf 22}^2$ transform in the
$(\bf{1},\bf{1})$ and $(\bf{2},\bf{2})$ representations of $SU(2)\times SU(2)$,
respectively. If one assumes that there is no inelastic diffractive
scattering~\cite{Weinberga}, there is an additional superconvergence relation which can be
expressed algebraically as

\begin{equation}
[ {\hat M}_{\bf 1}^2 , {\hat M}_{\bf 22}^2 ]\ =\ 0 \ .
\label{eq:msm3}
\end{equation}
This commutator constrains the mixing angles in reducible representations, and
has a peculiar interpretation as a discrete symmetry, as we will see below.

Consider the ground-state chiral multiplet for 
the $\lambda={1\over 2}$ baryons. The only representations of $SU(2)\otimes SU(2)$ that contain only
$I={1\over 2}$ and $I={3\over 2}$ states are
$({\bf 1},{\bf 2})$, $({\bf 2},{\bf 1})$, 
$({\bf 1},{\bf 4})$, $({\bf 4},{\bf 1})$, 
$({\bf 2},{\bf 3})$ and $({\bf 3},{\bf 2})$. One can easily convince oneself
that the unique reducible representation that contains the
$\lambda={1\over 2}$ $N$, $N^\prime$ and
$\Delta$, with no vanishing axial couplings or degeneracy among the states is
$({\bf 2},{\bf 3})\oplus ({\bf 1},{\bf 2})$~\footnote{
  Parity interchanges $SU(2)_L$ and $SU(2)_R$ representations. Therefore
  if we assign the 
  $\lambda=+{1\over 2}$ states to an $({\bf 2},{\bf 3})\oplus({\bf 1},{\bf 2})$
  representation, parity requires that the
  $\lambda=-{1\over 2}$ states are in the  $({\bf 3},{\bf 2})\oplus({\bf 2},{\bf 1})$
  representation~\cite{Weinberga}.}. This multiplet
was considered by Weinberg~\cite{Weinberga} (and also by Gilman and
Harari~\cite{Fred}). The
actual Dirac and Lorentz structure of the QCD interpolating fields that give
rise to this chiral structure is not at all
obvious. An early discussion of this problem can be found in work by
Casher and Susskind~\cite{CS74} and a recent discussion can be found in
Ref.~\cite{Beane2}. We will take the
$\lambda={3\over 2}$ $\Delta$ to transform as $({\bf 1},{\bf 4})$.

The helicity states of $N$
are in an $I={1\over 2}$ representation of
$SU(2)_I$, described by a tensor with a single fundamental index.  Likewise,
the helicity states of $\Delta$ are in an
$I={3\over 2}$ representation of $SU(2)_I$, 
described by a symmetric tensor with three
fundamental indices. To construct the $({\bf 2},{\bf 3})\oplus ({\bf 1},{\bf 2})$
representation which contains $N$, $N^\prime$ and $\Delta$,
we introduce the fields $S_a$, $T_{a,bc}$ to include the 
$\lambda=+{1\over 2}$ helicity states and the field 
$D_{abc}$ to include the $\lambda=+{3\over 2}$ helicity states.
The field $S_a$ 
transforms as $({\bf 1},{\bf 2})$ under $SU(2)_L\otimes SU(2)_R$; 
that is, $S\rightarrow LS$,
while the field $D_{abc}$ 
transforms as $({\bf 1},{\bf 4})$ under $SU(2)_L\otimes SU(2)_R$; 
that is,
$D\rightarrow LLL D$.
It is straightforward to embed an $I={1\over 2}$ and an $I={3\over 2}$ 
state into a single
irreducible representation of $SU(2)_L\otimes SU(2)_R$, the 
$({\bf 2},{\bf 3})$.
The field $T_{a,bc}$ transforms as $T\rightarrow RLLT$, and 
in terms of fields transforming as $I={1\over 2}$, $S_T$, 
and $I={3\over 2}$, $D_T$, $T$ can be written as
\begin{eqnarray}
T_{a,bc} & = & 
{1\over\sqrt{6}}\left(\ S_{T,b}\ \epsilon_{ac}\ +\ S_{T,c}\ \epsilon_{ab}\
\right)
\ +\ D_{T,abc} 
\ \ \ .
\end{eqnarray}
We also introduce a spurion field, $v^a_b$, which
transforms as $ v\ \rightarrow\  L\  v\  R^\dagger $, such that
$\langle {v^a_b} \rangle =M_{{\bf 2}{\bf 2} }^2\ \delta^a_b$.
The free-field dynamics of the helicity states
are determined by the two-dimensional effective Lagrange densities
constructed from the available tensors,
\begin{eqnarray}
{\mathcal L}_{\,\uparrow} & = &
\partial_+ T^{a,bc\dagger} \partial_- T_{a,bc}\ +\ 
\partial_+ S^{a\dagger} \partial_- S^{}_{a}
\ -\ M_{{\bf 1}T}^2\   T^{a,bc\dagger} T_{a,bc}
\ -\ M_{{\bf 1}S}^2\   S^{a\dagger} S^{}_{a}
\nonumber\\
& &\qquad\quad 
-\ {\mathcal A}\ \left(\  T^{a,bc\dagger} v^{d\dagger}_a S_{b} \epsilon_{cd}
\ +\ {\rm h.c.}\ \right) \ ,
\nonumber\\
{\mathcal L}_{\,\Uparrow} & = & \partial_+ D^{a,bc\dagger} \partial_- D_{a,bc}
\ -\ M_{{\bf 1}D}^2\   D^{a,bc\dagger} D_{a,bc}
\ ,
\end{eqnarray}
where ${\mathcal A}$ is an undetermined parameter 
and $x_{\pm} =z\pm t$ with $z$ the collinear direction. 
Notice that the helicity components of the baryons act as scalar
fields. The current operators
that satisfy the constraints imposed by 
Eq.~(\ref{eq:qcdsymm3}) take the form
\begin{eqnarray}
\hat T_\uparrow^\alpha 
& = & \ T^{a,bc\dagger} \left( T^\alpha \right)_a^{d} T_{d,bc}
\ + \ 2\  T^{a,bc\dagger} \left( T^\alpha \right)_b^{d} T_{a,dc}
\ +\  S^{a\dagger} \left( T^\alpha \right)_a^{d} S_d \ ,
\nonumber\\ 
\hat X_\uparrow^\alpha 
& = & \ T^{a,bc\dagger} \left( T^\alpha \right)_a^{d} T_{d,bc}
\ - \ 2\  T^{a,bc\dagger} \left( T^\alpha \right)_b^{d} T_{a,dc}
\ -\  S^{a\dagger} \left( T^\alpha \right)_a^{d} S_d \ ,\nonumber\\ 
\hat T_\Uparrow^\alpha 
& = & \ 3\ D^{abc\dagger} \left( T^\alpha \right)_a^{d} D_{dbc}\ ,
\nonumber\\ 
\hat X_\Uparrow^\alpha 
& = & \ -3\ D^{abc\dagger} \left( T^\alpha \right)_a^{d} D_{dbc}
\ \ .
\end{eqnarray}
The mass eigenstates are linear combinations of the chiral
eigenstates with a mixing angle $\psi$. Lorentz invariance requires
$M^2_{{\bf 1}T}=M^2_{{\bf 1}D}$ and one can easily check that the commutators of
Eq.~(\ref{eq:qcdsymm3}) and the mass-squared relation of Eq.~(\ref{eq:msm1}) are satisfied.
Diagonalizing the mass matrix and matching to the 
chiral perturbation theory current in Eq.~(\ref{eq:chiptaxial})
leads to
\begin{eqnarray}
& &g_A\ =\ 1\ +\ \textstyle{2\over 3}\cos^2\psi
\ \ ,\ \ 
g^\prime_A\ = \textstyle{2\over 3}\sin\psi\cos\psi
\ \ ,\ \ 
g^{\prime\prime}_A\ =\ 1\ +\ \textstyle{2\over 3}\sin^2\psi\ , \nonumber\\ 
& &\qquad{\mathcal C}_{\Delta N}\ =\ -2 \cos\psi
\ \ ,\ \ 
{\mathcal C}_{\Delta N^\prime}\ =\ -2 \sin\psi
\ \ ,\ \ 
{\mathcal H}_{\Delta\Delta}\ =\ -3 \ , \nonumber\\ 
& &\qquad\qquad  M_N^2\cos^2\psi
\ +\ M_{N^\prime}^2\sin^2\psi\  =\  M_\Delta^2 
\ ,
\label{eq:axials}
\end{eqnarray}
where $\psi$ is the mixing angle between the two $I={1\over 2}$ 
multiplets. One can readily verify that Eq.~(\ref{eq:axials}) parametrizes the complete set of A-W sum rules
for a pion scattering on $N$, $N^\prime$ and $\Delta$. In addition to
Eq.~(\ref{eq:AWpolesat}), the A-W sum rule for $\pi N$ scattering, one finds
\begin{eqnarray}
& &g_A^{\prime\prime\,2}+g_A^{\prime\,2}=1+{\textstyle{4\over 9}}\ {\mathcal C}_{\sss\Delta N^\prime}^2 \nonumber\\ 
& & {\mathcal C}_{\sss\Delta N}^2 +{\mathcal C}_{\sss\Delta
  N^\prime}^2={\textstyle{4\over 9}}\ {\mathcal H}_{\sss\Delta\Delta}^2 \nonumber\\ 
& &g_A g_A^{\prime}+g_A^{\prime}g_A^{\prime\prime}={\textstyle{4\over
    9}}\ {\mathcal C}_{\sss\Delta N}
{\mathcal C}_{\sss\Delta N^\prime}\nonumber\\ 
& &g_A {\mathcal C}_{\sss\Delta N}+g_A^{\prime}{\mathcal C}_{\sss\Delta  N^\prime}=
-{\textstyle{5\over 9}}\ {\mathcal C}_{\sss\Delta N}{\mathcal H}_{\Delta\Delta}\nonumber\\ 
& &g_A^\prime {\mathcal C}_{\sss\Delta N}+g_A^{\prime\prime}{\mathcal C}_{\sss\Delta  N^\prime}=
-{\textstyle{5\over 9}}\ {\mathcal C}_{\sss\Delta N^\prime}{\mathcal H}_{\Delta\Delta}
\ ,
\label{eq:otherA-Ws}
\end{eqnarray}
the A-W sum rules for $\pi N^\prime$ and $\pi \Delta$ scattering as well as the
inelastic A-W sum rules for the pairs $(N,N^\prime)$, $(N,\Delta)$ and $(N^\prime,\Delta)$, respectively.
\begin{table}[htb] 
\tbl{Axial couplings of the light-baryon ground-state chiral multiplet.
The third and second columns give the predictions of the 
$({\bf 2},{\bf 3})\oplus ({\bf 1},{\bf 2})$ representation both with (*) and without the
constraint of Eq.~(\ref{eq:msm3}).
The fourth column gives the axial couplings of the NCQM.
The experimental values have been determined via branching fractions that
appear in the particle data group~\protect\cite{pdg02}.
The extractions of ${\mathcal C}_{\Delta N}$ and
${\mathcal H}_{\Delta\Delta}$ from data were made in $SU(3)$ chiral perturbation theory~\protect\cite{BSS}.\vspace*{1pt}}
{\footnotesize
\begin{tabular}{||ccccc||}
    \hline
   &  $({\bf 2},{\bf 3})\oplus ({\bf 1},{\bf 2})$ & $({\bf 2},{\bf 3})\oplus ({\bf 1},{\bf 2})$* & NCQM  & EXPERIMENT  \\
\hline  \hline
\rule[-2mm]{0mm}{6mm} $|g_A|$  & $1 + \textstyle{2\over 3}\cos^2\psi$ &
${4\over 3}$  
& ${5\over 3}$ & $1.26$  \\
    \hline
\rule[-2mm]{0mm}{6mm} $|{\mathcal C}_{\Delta N}|$  & $2\cos\psi$ & $\sqrt{2}$ & $2$ & $1.2\pm0.1$ \\
    \hline
\rule[-2mm]{0mm}{6mm} $|{\mathcal H}_{\Delta\Delta}|$ & $3$ & $3$ & $3$ & $2.2\pm 0.6$  \\
    \hline
    \hline
\rule[-2mm]{0mm}{6mm} $|g_A^\prime|$  & $\textstyle{2\over 3}\sin\psi\cos\psi$ & $\textstyle{1\over 3}$ & $0$ & $0.71\pm0.20$  \\
    \hline
\rule[-2mm]{0mm}{6mm} $|{\mathcal C}_{\Delta N^\prime}|$  & $2\sin\psi $ & $\sqrt{2}$ & $0$ & $1.38\pm 0.50$  \\
    \hline
\end{tabular}\label{tab2} }
\vspace*{-13pt}
\end{table}
If we further impose the inelastic diffraction constraint of Eq.~(\ref{eq:msm3}), we find that
$M_{{\bf 1}T}^2 = M_{{\bf 1}S}^2$ and 
consequently $\psi={\pi\over  4}$, which
corresponds to maximal mixing. Notice that
this choice of the mixing angle corresponds to a discrete
symmetry of the free Lagrange density which interchanges $S$ and $S_T$~\cite{Beane2}.
This then gives
\begin{eqnarray}
& &\quad\qquad g_A\ =\ \textstyle{4\over 3}
\ \ \ ,\ \ \
g^\prime_A\ = \textstyle{1\over 3}
\ \ \ ,\ \ \
g^{\prime\prime}_A\ =\  \textstyle{4\over 3}  \ , \nonumber\\ 
& &{\mathcal C}_{\Delta N}\ =\ -\sqrt{2} 
\ \ ,\ \ 
{\mathcal C}_{\Delta N^\prime}\ =\ -\sqrt{2} 
\ \ ,\ \ 
{\mathcal H}_{\Delta\Delta}\ =\ -3 \ , \nonumber\\ 
& &\quad\qquad\qquad  M_\Delta^2\ -\ M_N^2\  \ =\ M_{N^\prime}^2\ -\ M_\Delta^2 \ .
\label{eq:axialswithdiff}
\end{eqnarray}
These values are impressively close to those in nature and it
is conceivable that the agreement may improve as the physical values are
extrapolated to the chiral limit. Using the nucleon and $\Delta$ masses as
input one finds $M_{N^\prime}=1467~{\rm MeV}$, consistent with the Roper
resonance~\cite{Ubi}. Notice that both $g_A$ and ${\mathcal C}_{\Delta N}$ are decreased from their NCQM
values ($\psi =0$) in the direction of experiment (see Table~\ref{tab2}).
Finally, we point out that the Roper-nucleon mass splitting is less than the chiral symmetry
breaking scale and therefore the non-vanishing quark-mass corrections to this chiral multiplet can
be computed using chiral perturbation theory~\cite{bsavage}. The convergence
of this chiral expansion is, of course, questionable. 

\section{The Naive Quark Model Interpretation}
\label{sec:NCQM}

\noindent The spin-flavor structure of the baryon multiplets seems to provide a
powerful explanation of why the A-W sum rule is almost completely
saturated by the $\Delta$, with smaller contributions from higher states. 
In the NCQM the nucleon and the $\Delta$ resonance fill out the
completely symmetric ${\bf 20}$-dimensional representation of spin-flavor
$SU(4)$, which we have seen is equivalent to the 
$(\bf{2} ,\bf{3} )$ representation of $SU(2)_L\times SU(2)_R$~\cite{Weinbergc,Beane4}.
In the NCQM the proton and $\Delta^+$ wavefunctions can be written as

\begin{eqnarray}
& & |{\;p\; ;\;{\bf 20}}>\ \sim {1\over\sqrt{6}}  (2|{u\uparrow u\uparrow d\downarrow}>
-|{u\uparrow u\downarrow d\uparrow}> - |{u\downarrow u\uparrow d\uparrow}> ) \ ,
\nonumber \\
& &|{\;\Delta^+\; ;\;{\bf 20}}>\ \sim {1\over\sqrt{3}}  (|{u\uparrow u\uparrow d\downarrow}>
+|{u\uparrow u\downarrow d\uparrow}> + |{u\downarrow u\uparrow d\uparrow}> )
\label{eq:QMint}
\end{eqnarray}
where cyclic permutations which are irrelevant for our purpose are not shown.
The action of the axial-vector operator, $q^\dagger \sigma^3\tau^3 q$, on
$u\uparrow$ and $d\downarrow$ is $+1$ and on
$u\downarrow$ and $d\uparrow$ is $-1$. One then trivially finds $\ga =5/3$ and
${\mathcal C}_{\sss\Delta N}=2$. Similarly, placing the proton in the completely antisymmetric
${\bf 4}$-dimensional representation gives rise to the wavefunction

\begin{equation}
|{\;p\; ;\;{\bf 4}}>\ \sim {1\over\sqrt{2}} (|{u\uparrow u\downarrow d\uparrow}>
-|{u\downarrow u\uparrow d\uparrow}>)\; +\ldots
\label{eq:QMint1}
\end{equation}
from which one finds $\ga =1$. We then recover the
axial-coupling predictions from our minimal realistic model by placing $N$, $N'$
and $\Delta$ in a reducible ${\bf 4}\oplus{\bf 20}$ representation of
$SU(4)$:

\begin{eqnarray}
& & |{N}>=\sin\theta\,|{\;{\bf 4}\; ;\;
1^+}>+\cos\theta\,|{\;{\bf 20}\; ;0^+}>_{4},
\nonumber \\
& &|{N'}>=-\cos\theta\,|{\;{\bf 4}\; ;\; 1^+}>+\sin\theta\,|{\;{\bf 20}\;
;0^+}>_{4},
\nonumber \\
& & |{\Delta}>=|{\;{\bf 20}\; ;0^+}>_{16} \ ,
\end{eqnarray}
where the subscripts indicate the spin-flavor content.
Here we have included the spatial quantum numbers that one naively expects. Since the
${\bf 4}$ of spin-flavor $SU(4)$ is completely antisymmetric, it must carry at least one
unit of orbital angular momentum. In the NCQM the $|{\;{\bf 4}\; ;\; 1^+}>$
state ($|{\;{\bf 20}\; ;\; 1^+}>$ of $SU(6)$) is thought to be irrelevant as it requires two quarks in a baryon
to be in an excited state. The presence of orbital angular momentum is quite
strange as a nonvanishing nucleon-$\Delta$ mass splitting requires that
${\hat M}_{\bf 22}^2$, which acts like an order parameter, carry
orbital angular momentum. The peculiar NCQM interpretation of the chiral
symmetry representations in the collinear frame was noticed long ago by
Casher and Susskind~\cite{CS74}.

In the NCQM one usually assigns $N'$ and $\Delta'$ to a radially-excited
${\bf 20}$-dimensional representation of $SU(4)$. These states 
then mix with the ``ground state'' ${\bf 20}$-dimensional representation containing
$N$ and $\Delta$. This reducible model has been analyzed in Ref.~\cite{Ubi};
it overpredicts the Roper mass and has no solution for the axial couplings
consistent with the expected chiral transformation properties of the
mass-squared matrix.

\section{Conclusions}
\label{sec:conc}

\noindent The main point to take from this talk is that even at low energies
where chiral symmetry is spontaneously broken, there is a sense in which the
baryon spectrum falls into {\it reducible} representations of the chiral
algebra. This has nothing to do with parity doubling near a
chiral symmetry restoring phase transition. We have found that existing data
suggest that the nucleon and the $\Delta$ and Roper resonances form a reducible
sum of $(\bf{1}, \bf{2})$ and $(\bf{2} ,\bf{3})$ representations of the chiral group,
with maximal mixing. From the perspective of the naive constituent quark model
this is equivalent to placing these states in a reducible 
${\bf 4}\oplus{\bf 20}$ representation of
spin-flavor $SU(4)$. Our results suggest that other baryons also fall into
finite-dimensional chiral representations that in principle can be mapped out
at JLab and other experimental facilities. Of course
as one moves higher in the excited spectrum one expects that the assumption
of pole saturation becomes increasingly unreliable due to the broadening
of hadronic states. We stress that it is somewhat peculiar that the chiral multiplet involving the
nucleon involves only a few states and that the representations enter with
approximately equal weight~\cite{Weinbergd,Beane2,SavageBeane}. Recently, we
have conjectured that the ground-state light baryons, the heavy baryons, and the
light and heavy mesons fall into chiral multiplets which
contain the minimal particle content necessary to saturate the interpolating
fields for the hadrons, and which allow for non-zero mass splittings between
members of the multiplet~\cite{SavageBeane}.  The claim that chiral multiplets are small and
decoupled might appear odd given that there are observed axial transitions
---say in the light-baryon sector--- from excited states to the ground-state
multiplet.  However, the decoupled chiral multiplets mix when the quark masses
are turned on. Therefore, the smallness of the axial transitions from excited
multiplets to the ground-state multiplet as compared to those within the
ground-state multiplet is conjectured to be due to the smallness of the quark masses.
Improved data both from experiments and from lattice QCD, combined with 
theoretical work assessing chiral corrections will ultimately either confirm or
refute this conjecture. The structure of the chiral
representations filled out by the hadrons remains an important unsolved problem in QCD.

\section*{Acknowledgments}
I thank Martin Savage and Bira van Kolck for enjoyable collaborations
and Eric Swanson and Steve Dytman for an enjoyable workshop.
This work is supported in part by the U.S. Department of Energy under 
Grant No.~DE-FG03-00-ER-41132.

\end{document}